%% file: main.tex
\documentclass{article}
\pdfoutput=1

\usepackage{arxiv}
\usepackage[utf8]{inputenc}
\usepackage[T1]{fontenc}
\usepackage{hyperref}
\usepackage{url}
\usepackage{amsfonts}

\usepackage{siunitx}
\usepackage{subcaption}
\usepackage{multirow}
\usepackage{xcolor}
\usepackage{graphicx}
\usepackage{booktabs}

\definecolor{mygreen}{RGB}{80, 200, 50}
\definecolor{myred}{RGB}{230, 30, 30}
\newcommand{\cfbox}[2]{%
\colorlet{currentcolor}{.}%
{\color{#1}%
\fbox{\color{currentcolor}#2}}%
}

\title{Spectrally Consistent Unet for High Fidelity Image Transformations}

\author{
    Demetris Marnerides\\
    University of Warwick, UK\\
    \texttt{Demetris.Marnerides@warwick.ac.uk}\\
    \And
    Thomas Bashford-Rogers\\
    University of the West of England, UK\\
    \texttt{Tom.Bashford-Rogers@uwe.ac.uk}\\
    \And
    Kurt Debattista\\
    University of Warwick, UK\\
    \texttt{K.Debattista@warwick.ac.uk}\\
}

\begin{document}
\maketitle

\begin{abstract}
\input{abstract.tex}
\end{abstract}

\input{introduction.tex}
\input{background.tex}
\input{guided_unet.tex}
\input{spectra.tex}

\input{itm.tex}

\input{colourisation.tex}
\input{conclusion.tex}

\bibliographystyle{unsrt}
\bibliography{ref}

\end{document}

%% file: abstract.tex
Convolutional Neural Networks (CNNs) are the current de-facto models used for many imaging tasks due to their high learning capacity as well as their architectural qualities. The ubiquitous UNet architecture provides an efficient and multi-scale solution that combines local and global information. Despite the success of UNet architectures, the use of upsampling layers can cause artefacts. In this work, a method for assessing the structural biases of UNets and the effects these have on the outputs is presented, characterising their impact in the Fourier domain. A new upsampling module is proposed, based on a novel use of the Guided Image Filter, that provides spectrally consistent outputs when used in a UNet architecture, forming the Guided UNet (GUNet). 
The GUNet architecture is applied  and evaluated for example applications of inverse tone mapping/dynamic range expansion and colourisation from grey-scale images and is shown to provide higher fidelity outputs.

%% file: introduction.tex
\section{Introduction}\label{sec:introduction}

Image transformation problems can be addressed using end-to-end training of Convolutional Neural Networks (CNNs). Such problems include colourisation~\cite{iizuka2016colornet}, super-resolution~\cite{sugawarasr} and inverse tone mapping~\cite{eilertsen2017cnn}. Solutions to these problems are required to be multi-scale, combining spatially local and global information, but must also be power and memory efficient. One of the most popular CNN architectures for such problems is the ubiquitous UNet~\cite{ronneberger2015unet}, which has been used extensively for image transformation problems~\cite{isola2016pix2pix,eilertsen2017cnn,endo2017drtmo,jin2017inverse}. Despite the broadly positive results achieved using UNet architectures, it has been noted on multiple occasions \cite{pix2pixsuppl, zhang2017suppl, sugawarasr, marnerides2018exp} that the upsampling layers can cause artefacts to appear in the output, especially via the use of the commonly-implemented transposed convolutional layers which cause checkerboard-like patterns~\cite{odena2016deconvolution}.

In order to resolve such issues, a novel module, which generalises guided image filtering (GIF)~\cite{hegif}, is introduced to replace the standard upsampling and concatenation modules of the UNet architecture. The module is used within a UNet architecture to form what shall be termed a Guided UNet (GUNet).

A spectral analysis is proposed to investigate the structural properties of CNNs in the frequency domain. The method compares the spectra of the outputs of multiple network configurations, showing the effects that the commonly used upsampling modules cause on the outputs and the pre-existing biases that UNets have architecturally. 

The proposed GUNet architecture alleviates the effects shown in the spectral investigation, minimising the structural biases of traditional UNets through guided feature upsampling, which preserves the spectrum of the input. In particular,
the proposed architecture diminishes the checkerboard artefacts and/or frequency suppression that arises from using upsampling layers.
Examples of such artefacts can be seen in Figure~\ref{fig:spectra_model_avg}, Figure~\ref{fig:big_samples} and Figure~\ref{fig:samples}.
While these artefacts are kept to a minimum, the efficiency of UNet architectures is maintained both in memory and computational speed. In order to demonstrate the potential of the method, GUNet is used to solve the inverse tone mapping (ITM) problem~\cite{banterle06itm} and results show that it compares favourably with state-of-the-art methods~\cite{eilertsen2017cnn, marnerides2018exp}. Colourisation is also presented, showing improvements when using GUNet compared to alternative UNet architectures.

In summary, the main contributions of this work are:
\begin{itemize}
    \item A new feature upsampling module that improves the output image quality produced by UNet architectures.
    \item A novel spectral investigation of the properties of UNet architectures, demonstrating the advantages of GUNet.
    \item Trained GUNet architectures showcasing results for ITM and colourisation as example applications, with state of the art performance for ITM.
\end{itemize}

%% file: background.tex
\section{Background and Related Work}\label{sec:background}

\input{unet_diagrams_only_gunet}

The UNet architecture is based on autoencoder networks ~\cite{schmidhuber2014deep} and uses downsampling and subsequently upsampling layers for improved efficiency and expressiveness.  The bulk of the computation is performed on lower resolutions and is thus faster and uses less memory.  The encoder part of the architecture uses downsampling and convolutional layers to produce a low resolution encoding of the input, which is passed through a bottleneck and is then upsampled and processed by the decoder to produce the final output, similarly to Figure~\ref{fig:unet_diagrams}.

Downsampling can be conducted using a variety of methods, including max-pooling, average-pooling or strided convolutions. 

The lower resolution feature maps can be upsampled back to full resolution by the use of upsampling layers. Upsampling layers include learnable transposed convolutions~\cite{longfullyconv}, or more traditional upsampling algorithms like nearest neighbour or bilinear interpolation. Transposed convolutional layers are computationally equivalent to the operations performed when normal strided convolutional layers are backpropagated for the gradient computation.

Each layer of convolutions and downsampling in the architecture inevitably acts as a low pass filter, suppressing details and higher spatial frequencies present in the input\footnote{This is unless the convolution filter only depends on the central pixel (i.e.\ all other filter values are zero), or it is of size \(1 \times 1\), which does not alter the receptive field.}. For the higher frequencies to be propagated to the output, the encoder needs to learn to encode them in the intermediate features of the network.

Ronneberger, Fischer and Brox~\cite{ronneberger2015unet} introduced skip-connections to the autoencoder architecture to form the UNet architecture. Skip-connections concatenate the encoder with the decoder features at each level of the decoder, bypassing the lower levels of the network. This helps to better propagate details from the input to the output without having to learn to encode them, thus allowing for the lower levels to better encode more global features.  UNet architectures, are good at combining local and global scales since the receptive field of their lower levels is quite large and is progressively upsampled and combined with the more local scales in the decoder.  

However, the use of upsampling layers in the decoder can cause checkerboard artefacts or blurring~\cite{odena2016deconvolution}, depending on the type of layer used and the content it is applied on. Transposed convolutions, or the mathematically equivalent faster implementation termed Sub-Pixel convolution~\cite{shi2016real}, are prone to checkerboard artefacts~\cite{pix2pixsuppl,zhang2017suppl}, while non-learned upsampling methods can cause blurring, since they are based on pre-defined interpolation. In addition, upsampling layers cause information bleeding in low contrast areas, particularly ones close to sharp boundaries. Skip connections help alleviate some of these problems but are not sufficient, as artefacts can be observed in multiple cases of fully trained UNet architectures~\cite{pix2pixsuppl, zhang2017suppl, sugawarasr, marnerides2018exp}.

A number of methods have been presented that attempt to alleviate upsampling artefacts present in UNet architectures. Odena, Dumoulin and Olah~\cite{odena2016deconvolution} propose the use of ``resize'' convolutions, where the features are first upsampled using nearest-neighbour or bilinear upsampling followed by convolution. Wojna et al.~\cite{wojna2017devil} study variants of such configurations. Aitken et al.~\cite{aitken2017checkerboard} propose a specialised initialisation scheme for transposed convolutional layers that correlates the kernel weights at initialisation. Sugawara, Shiota and Kiya~\cite{sugawarasr} propose a similar approach for use in super-resolution, which however correlates the kernel weights within the architecture and not at initialisation.

While these methods may alleviate artefacts, they do not effectively use the highly detailed information which already exists in the encoder, but rather try to recover it in the decoder. This approach can be prone to blurring, since the methods are based on pre-defined interpolation or correlation and do not provide robust results that preserve the input structure accurately. In addition, learned upsampling using transposed convolutions, even with a correlating initialisation and a carefully chosen loss function, is not guaranteed to avoid artefact producing minima after optimisation. Using the proposed guided upsampling module,
which incorporates higher level encoder features as the guidance ``image'', high frequencies can be successfully transferred to the upsampled features of the decoder. This is due to the edge-preserving nature of the proposed guided filtering module at each level.

%% file: unet_diagrams_only_gunet.tex
\begin{figure}[t!]
\centering
\includegraphics[width=1.0\linewidth]{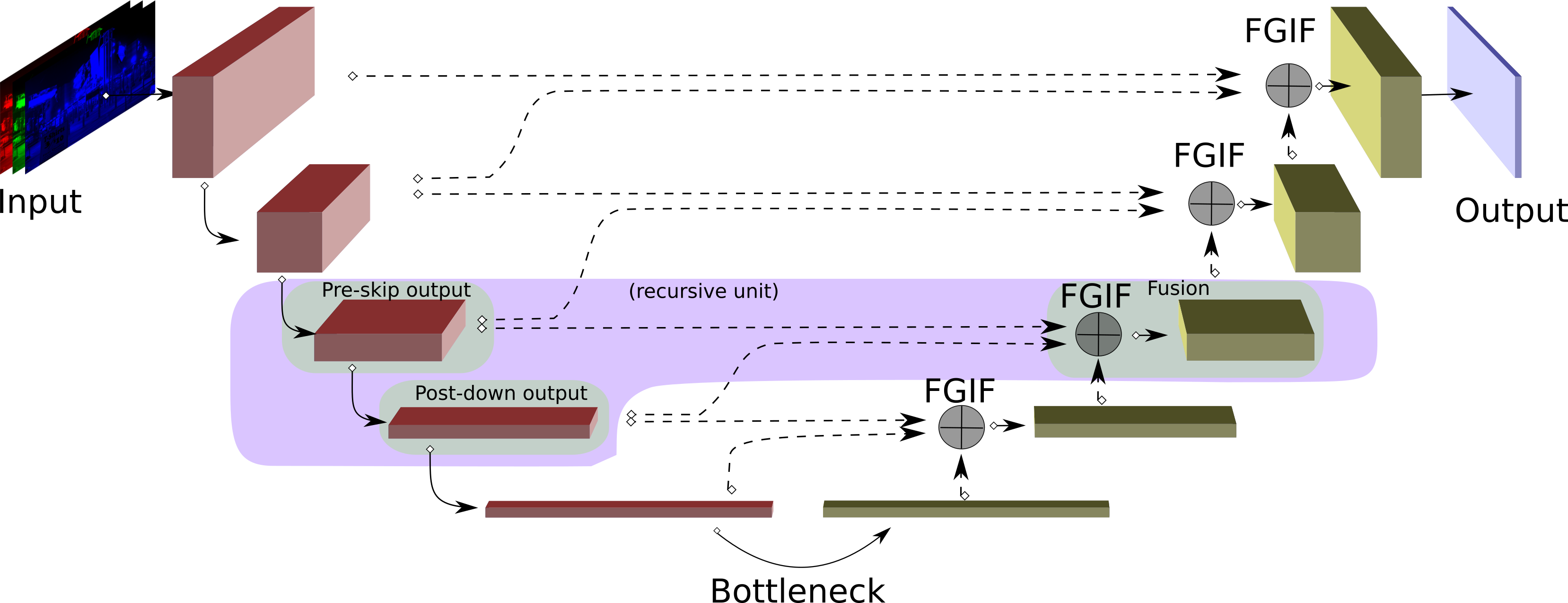}
\caption{The Guided UNet architecture.}\label{fig:unet_diagrams}
\end{figure}

%% file: guided_unet.tex
\section{GUNet: Guided UNet}\label{sec:gunet:model}

This section introduces the GUNet model in which the upsampling and concatenation modules of the UNet architecture are replaced to improve fidelity.
The proposed replacement module is able to upsample the decoder features, and simultaneously transfer high frequency information from the encoder.
A variety of imaging techniques have been introduced that can transfer detail from one image to another, for example the joint bilateral filter~\cite{kopf2007joint} or the GIF~\cite{hegif}.

Most importantly, these techniques can leverage higher resolution guides, to not only transfer detail but to also upsample at the same time using the high resolution image as a guide. Thus, the proposed module uses the fast GIF~\cite{hefastgif} to jointly filter and upsample the decoder features, using the corresponding features in the encoder.

\subsection{Guided Feature Upsampling}\label{sec:gunet:gif_module}

The proposed network makes use of a new module based on the GIF~\cite{hegif}. The GIF is a differentiable edge preserving filter that can also be used for guided upsampling~\cite{hefastgif}. When used in a decoder-encoder architecture, to filter a decoder feature \(z\) using the encoder feature \(y\) as a guidance, the resulting feature \(q\) is assumed to be a locally linear model of the guidance \(y\), similarly to the GIF. For a spatial feature neighbourhood \(\omega_k\) of (square for simplicity) size \(n\), containing \(N = n^2\) pixels:

\begin{equation}
q_i = \bar{a}_k y_i + \bar{b}_k, \forall i \in \omega_k
\end{equation}

\noindent with the constants (in \(\omega_k\)) \(\bar{a}_k\) and \(\bar{b}_k\) given approximately by using linear ridge regression by:

\begin{equation}\label{eqn:gif_a}
 \bar{a}_k = \frac{1}{N}\sum_{k \in \omega_i}
\frac{\frac{1}{N}\sum_{i \in \omega_k}y_iz_i -\mu_k\bar{z}_k}{\sigma_{k}^{2}+\epsilon}
\end{equation}
\begin{equation}\label{eqn:gif_b}
 \bar{b}_k = \frac{1}{N}\sum_{k \in \omega_i} \bar{z}_k -
 \bar{a}_k\mu_k
\end{equation}

\noindent where \(\mu_k\) and \(\sigma_{k}^2\) are the mean and variance of the guidance feature in \(\omega_k\). \(\epsilon\) is a regularisation parameter which penalises the effects of the guidance, by adjusting the value of \(\bar{a}_k\).  This approximation does not guarantee that \(\bar{a}_k\) and \(\bar{b}_k\) are constants in \(\omega_k\), however they still preserve large gradients (strong edges) from the encoder guidance feature, since:

\begin{equation}\label{eq:gunet:edge}
\nabla q \approx \bar{a}_k\nabla y.
\end{equation}

Similarly to the fast GIF implementation~\cite{hefastgif}, which can be used for guided upsampling, \(y\) and \(z\) are the corresponding downsampled features of the architecture and are used to compute \(\bar{a}_k^{\text{lr}}\) and \(\bar{b}_k^{\text{lr}}\) on the lower resolution.  The coefficients are then upsampled back to the higher resolution to form \(\bar{a}_k^{\text{hr}}\) and \(\bar{b}_k^{\text{hr}}\), using bilinear upsampling. The coefficients are then applied on the higher resolution guidance feature \(x\) to compute the final filtered decoder feature:

\begin{equation}\label{eqn:gif_hr}
q_i = \bar{a}_k^{\text{hr}} x_i + \bar{b}_k^{\text{hr}}.
\end{equation}

Guided feature upsampling combines the encoder and decoder features of the architecture and aims to guide its features at each upsampling stage in the output to be structurally similar to the corresponding feature set of the input features in the encoder.

Wu et al.~\cite{wutrainablegif} use the GIF in conjunction with deep learning, introducing the deep guided filter and a derivation of an analytic form of the derivative of the filter, which is useful for implementing backpropagation for the GIF without using automatic differentiation software. However, in the case of the deep guided filter, a network is used within a guided filter to model the mapping from the guide to the input, at a lower resolution and is trained end-to-end along with the filter from scratch. In the case of the GUNet architecture the opposite is proposed, where the guided feature upsampling module is used within a UNet architecture, as many times as needed, to improve the decoder fidelity.

\input{gunet_slice_diagram}

\subsection{Model Architecture}

Figure~\ref{fig:unet_diagrams} shows a UNet architecture with a highlighted slice at a specific level, which can be thought of as the basic recursive element of the architecture.  Figure~\ref{fig:gunet} shows a generalised version of the recursive element in detail. The input at that level is first ``pre-processed'', with the resulting features \(x\) downsampled to form a lower resolution set \(y\).  \(y\) is used as the input to a child level of the same form, or the lowermost bottleneck module. The output of the child module, \(z\) can then be fused with \(x\) and \(y\) in one of the ways depicted in the purple boxes. The Pre-Skip, Down and Post-Down components are part of the encoder, while the fuse and post-fuse components are part of the decoder.

In the case of autoencoders, \(z\) is upsampled and is not combined with \(x\) and \(y\), thus relying only on the information encoded in the bottleneck, which might be useful in some applications, for example compression. The UNet architecture performs the same upsampling, but also combines the features \(x\) with the upsampled \(z\) and fuses them using a convolutional layer, usually of kernel size \(1 \times 1\).  The network also uses residual connections at various points to better propagate gradients. The specific configurations with regards to the ordering of the normalisation layers, activations and convolutions are adapted from the proposed sequence described by He et al.~\cite{heidentityresnet}.

The GUNet architecture replaces the upsampling/fusion layer with the proposed guided filtering module.  In this case, the features, \(x\), serve as the high resolution guidance image, while the lower resolution features \(z\) are the filter input. The filter is applied separately on each feature channel. The low resolution input, \(y\), and child output \(z\) are used as the guidance and input images, in Equation~\ref{eqn:gif_a} and Equation~\ref{eqn:gif_b}, respectively, to compute \(\bar{a}_k^{\text{lr}}\) and \(\bar{b}_k^{\text{lr}}\). These coefficients are then upsampled using bilinear upsampling and applied on the guidance features \(x\) using Equation~\ref{eqn:gif_hr}.

It is worth pointing out that GUNet results in fewer parameters than UNets with transposed convolutions or bilinear upsampling, since the upsampling is parameter free and the concatenation layer is avoided.

%% file: gunet_slice_diagram.tex
\begin{figure*}[t]
    \centering
    \includegraphics[width=1.0\linewidth]{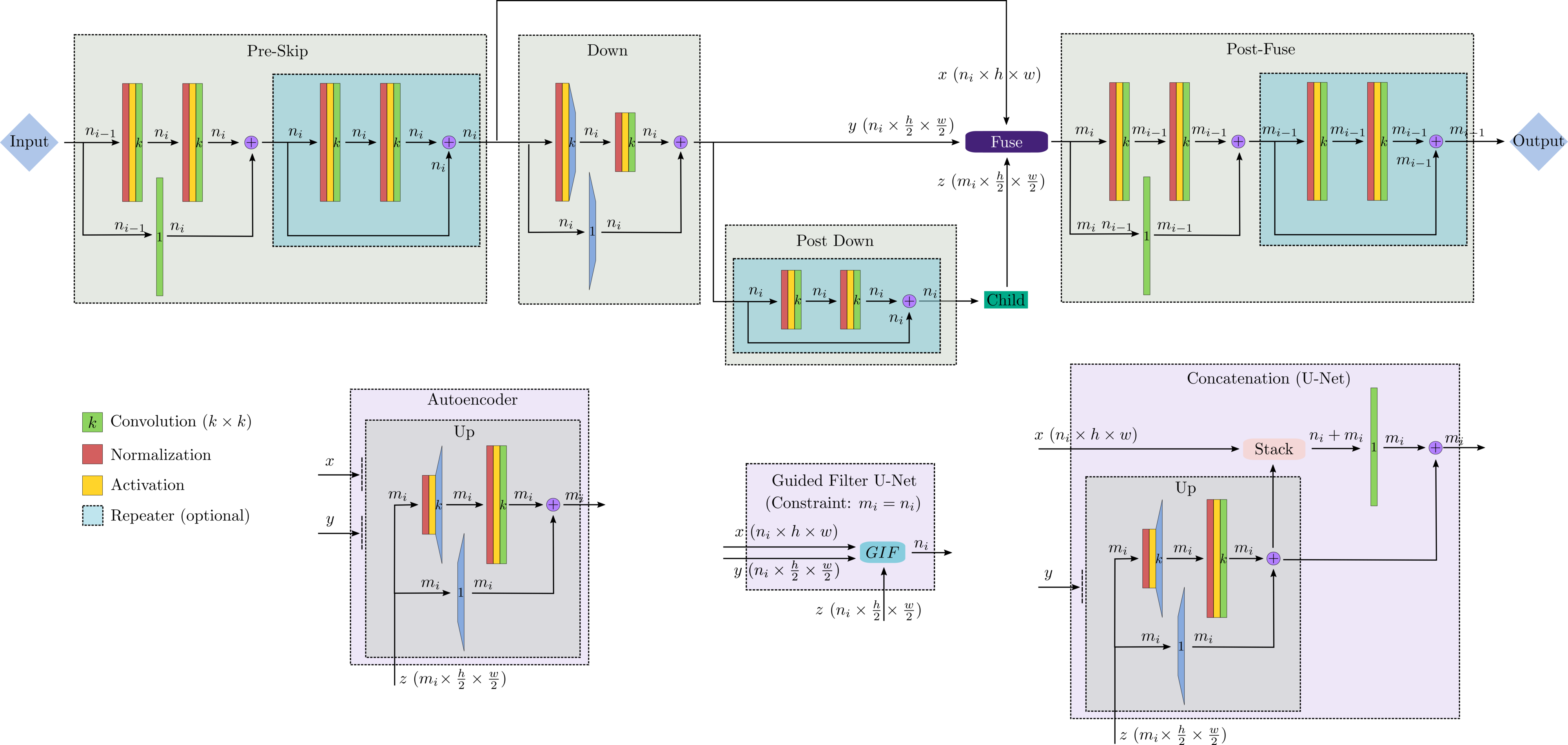}
    \caption{Slice of a
        single level from a UNet. The purple fusion
         blocks determine whether the network is an Autoencoder,
        a UNet or a GUNet. The proposed module uses the fast guided filter to
guide feature upsampling in the decoder.}\label{fig:gunet}
\end{figure*}

%% file: spectra.tex
\section{Spectral Analysis \& Results}\label{sec:gunet:spectra}

This section presents a method for analysing the effect that upsampling layers have on UNet network predictions. The proposed spectral investigation aims to identify and explain the sources of artefacts in UNet architectures. 
Due to the regularly repeating nature of the checkerboard artefacts, it is hypothesised that such artefacts will consistently alter the output image spectrum due to the introduction or suppression of specific spatial frequencies. This can help identify and compare the effect of different modules but also provide a way to judge the structural properties of any alternative proposals.

\subsection{Method}\label{sec:gunet:spectra_method}

An image can be decomposed into two-dimensional discrete spatial waves of tone variation whose weighted combination (spectrum) fully characterise it. The image spectrum can be computed using the discrete 2D Fourier transform and is composed of complex values, consisting of a phase and magnitude. For the purposes of this work, the main focus will be on the spectrum magnitude
which is more interpretable than the phase for describing spatial artefacts. The three channels of coloured RGB images are calculated separately and are averaged. The magnitude of the spectrum is radial from the centre, with high frequencies being more central. Usually, high frequencies are depicted on the boundaries, but for better visualisation of the comparisons they are depicted in the centre.
The brighter the pixel, the higher the magnitude of the corresponding frequency is in the original image.

The spectrum of the outputs of multiple networks is computed and used as an evaluation of the structural bias of the underlying model architecture. Specifically, the structural properties of the upsampling modules used in UNet architectures can be investigated by observing their effects on the spectrum of the output images. 
The upsampling modules under consideration are the transposed convolution (TC), nearest neighbour (NN) and bilinear interpolation (BI), which are the most commonly used in CNNs~\cite{odena2016deconvolution}. Three UNet architectures TC-Unet, NN-Unet and BI-Unet along with a GUNet architecture, are presented and compared. 

The architectures follow the design from Figure~\ref{fig:gunet}. The encoder for all architectures downsamples four times, similarly to the original UNet architecture~\cite{ronneberger2015unet}, with feature sizes 16, 32, 64 and 128, matched by the decoder. A kernel size of \(3 \times 3\) is used except in the Pre-Skip and Post-Fuse modules which use \(1 \times 1\) convolutions, such that the detail is not inadvertently filtered at those points. For TC-Unet, the transpose convolutions are of kernel size \(4 \times 4\). This is to avoid any overlap issues that occur when combining stride-two convolutions with odd-sized kernels as described by Odena, Dumoulin and Olah \cite{odena2016deconvolution}. The ReLU activation is used along with batch normalisation, as is the current de-facto standard for CNNs. The bottleneck consists of four residual blocks of 128 features each containing two convolutional layers, exactly the same as a Post-Down module from Figure~\ref{fig:gunet} with four repeater units. There are no repeater units in any of the other modules.

\input{spectra_comparisons}

\subsection{Results}\label{sec:gunet:spectra_results}

Figure~\ref{fig:spectra_model_avg} shows the spectra of the inputs and the corresponding outputs for the four architectures. Each architecture's parameters are sampled 50 times using the gaussian distribution initialisation described by He et al.~\cite{he2015delving}, which is best suited for ReLU activations. The outputs of all the samples of each architecture are averaged in this manner, in order to marginalise out any weight initialisation biases in the comparisons. The top two rows show an example from a single input image. The first row depicts the average of the outputs and the input in the spatial domain, while the second row shows the corresponding magnitudes of the averages of the spectra of the output images. Exposures are taken for better visualisation of the very high dynamic range of the spectrum. The third row shows the average spectra of the outputs from all models and their inputs, averaged over 50 model samples and also over 50 different input images.

Column (a) shows the result from TC-UNet. The checkerboard artefacts in the output image are a result of the transposed convolution and are translated into regular peaks of dominant frequencies in the Fourier domain.  The central pixel in the spectrum image is one of the brightest and it corresponds to the smallest checkerboard patterns (2-pixel period) which are clearly visible in the output. Column (b) shows results for the NN-UNet configuration, where the transposed convolutions are replaced with nearest neighbour for upsampling followed by a convolution. In this case, the output tends to be more blurry which is reflected in the output spectrum where there are patterns of higher frequencies being suppressed thus appearing darker in the spectrum. The BI-Net configuration results are shown in column (c). Higher frequencies are suppressed (the overall spectrum slices are darker than the input spectrum towards the centre) but there is much improvement compared to the effects of the transposed convolution and nearest neighbour upsampling. Column (d) shows results using a GUNet architecture. There are no apparent artefacts in the spectrum, which is mostly preserved, with the higher frequencies not appearing darker or distorted compared to the other architectures.

\subsection{Discussion}\label{sec:gunet:spectra_discussion}

The UNet models presented above exhibit persistent artefacts due to their architectures. Transpose convolutions in the decoder introduce high frequencies and favour some over the others, while nearest neighbour upsampling suppresses high frequencies. Bilinear upsampling produces better output spectra compared to nearest neighbour which severely suppresses specific frequencies.
Models that contain these modules are structurally biased towards producing artefacts. These effects can possibly be diminished by constructing training losses that direct the network weights to counteract these structural biases. This can be hard (or impossible) for pixel-wise losses, for example the \(L_1\) or \(L_2\) norms, since they do not take into account inter-pixel correlations which will inform the training procedure with respect to the spectrum.

Such networks have a weight configuration space highly populated with artefact producing points, that either promote or suppress specific frequencies.  This does not mean that a subset of non artefact producing points (sets of weights) does not exist, nor that such a set is not reachable after sufficient training/fine-tuning or by using a loss which specifically aims to do so. However there are no good reasons to select an artefact-biased architecture to begin with. On the contrary, a less biased architecture can lead to improved results, since it must not un-learn existing biases. The GUNet architecture is specifically designed to avoid such biases while maintaining the benefits of the traditional UNet architecture as can be seen in these results.

%% file: spectra_comparisons.tex
\begin{figure*}[t!]
    \centering
    \begin{subfigure}{0.17\linewidth}
        \includegraphics[width=1.0\linewidth]{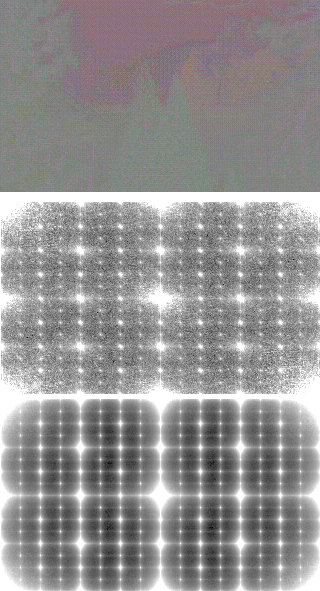}
        \caption{TC-UNet}
    \end{subfigure}
    \begin{subfigure}{0.17\linewidth}
        \includegraphics[width=1.0\linewidth]{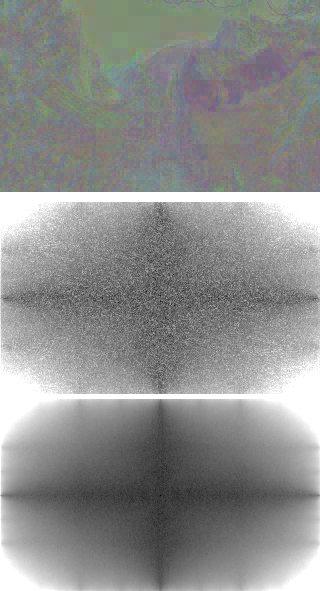}
        \caption{NN-UNet}
    \end{subfigure}
    \begin{subfigure}{0.17\linewidth}
        \includegraphics[width=1.0\linewidth]{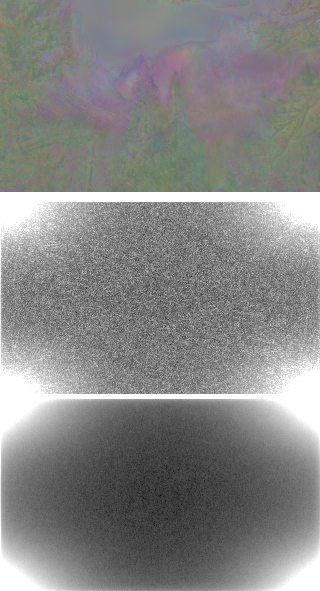}
        \caption{BI-UNet}
    \end{subfigure}
    \begin{subfigure}{0.17\linewidth}
        \includegraphics[width=1.0\linewidth]{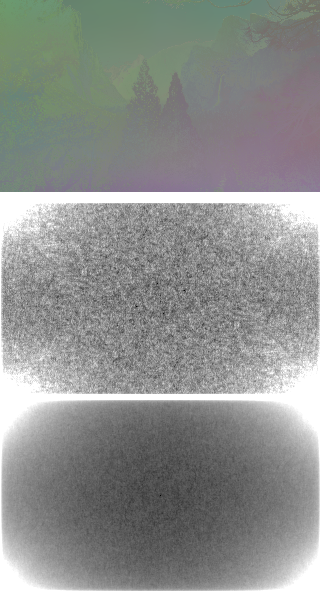}
        \caption{GUNet}
    \end{subfigure}
    \begin{subfigure}{0.17\linewidth}
        \includegraphics[width=1.0\linewidth]{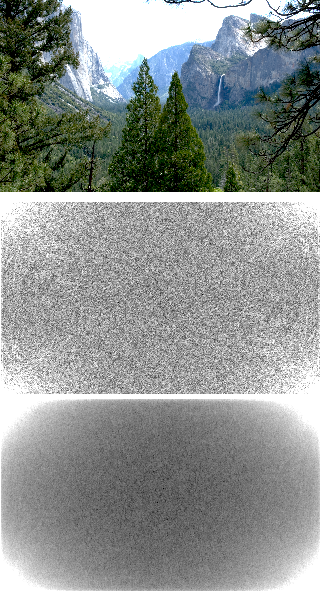}
        \caption{Input}
    \end{subfigure}\\
    \caption{Comparison of network outputs (top) and their spectra (middle), averaged over 50 network initializations. The bottom row shows spectra averaged over 50 different inputs (each initialised 50 times).
    }\label{fig:spectra_model_avg}
\end{figure*}

%% file: itm.tex
\section{Application: Inverse tone mapping}\label{sec:application}

\input{bigsample.tex}
\input{samples}
\input{tables}

This section demonstrates the use of GUNet for a real-world problem, inverse tone mapping (ITM) also known as dynamic range expansion. ITM is the problem of recovering High Dynamic Range (HDR) from a standard or Low Dynamic Range (LDR) image. ITM is a robust test for assessing the fidelity of image transformation networks due to its extreme contrast in the output and its inverse nature. A number of methods to do this exist \cite{banterle06itm, akyuz2007hdr}, however, recently these have been superseded by deep learning solutions, some based on UNets \cite{eilertsen2017cnn} and others that are more dedicated \cite{marnerides2018exp}.
The architecture used is the same as the GUNet model described in the spectral analysis section. The guided feature upsampling modules use \(\epsilon = 0.001\). Lower epsilon values lead to stronger guidance, which is necessary for this problem, since the gradient structure needs to be preserved. The GIF kernel size is chosen to match the full width and height of the feature map at each level. It is worth noting that the size of the filter adapts to the size of the inputs at inference time to adjust for different image sizes.

\input{colour_figure}

We follow the training and evaluation procedure presented in ExpandNet~\cite{marnerides2018exp}. The training dataset consists of 1,013 HDR images of different resolutions and the testing dataset consists of 50 test images from the Fairchild Photographic Survey~\cite{fairchild2007hdr}.The LDR inputs are generated on-the-fly during training using four randomised tone mapping operators and exposures. The Adam optimiser is used, with default parameters and a learning rate of 3e-4 and a batch size of \(32\). The input LDR images are mapped to the \([0, 1]\) range. The loss optimised is the L1 and cosine similarity with \(\lambda=5\). The network is trained for approximately one week for a total of 720,000 iterations using an Nvidia 2070 SUPER GPU\@ using PyTorch~\cite{pytorch}.

\subsection{Results}\label{sec:gunet:results}

Quantitative comparisons between the trained GUNet (GUNet) and the state-of-the-art ITM methods by 

Eilertsen et al.~\cite{eilertsen2017cnn} (EIL) and Marnerides et al.~\cite{marnerides2018exp} (EXP) are presented, along with a UNet architecture with transposed convolution layers for upsampling (TC-UNet) and a UNet with bilinear interpolation upsampling (BI-UNet). The evaluation method 
uses the perceptually uniform (PU)~\cite{aydin2008extending} encoded metrics, SSIM~\cite{wang2004image}, MS-SSIM~\cite{wang2003multiscale} and HDR-VDP-2~\cite{narwaria2015hdr}, traditionally used for evaluating HDR images, for both \textit{optimal} and \textit{culling} (clipping of top and bottom \(10\%\) of values) exposures, and the \textit{scene-referred} (scaling to original HDR image range) and \textit{display-referred} (scaling to 1000 nits display range) settings. The PU encoding accounts for the non-linear response of the human visual system to luminance and adapts traditional LDR metrics for HDR. 

Table~\ref{table:gunet_results} shows the average test performance for all metrics and scenarios.
GUNet performs well, achieving the highest values in the optimal exposure setting. This is in line with its design, which relies on existing information regarding the structure and edges of the images to guide the dynamic range expansion in the result.
To better showcase the importance of the architecture and the guidance, example images are presented for predictions from TC-UNet, BI-UNet, EXP and GUNet in Figure~\ref{fig:big_samples} and Figure~\ref{fig:samples}. The predicted HDR images produced using TC-UNet, BI-UNet and EXP exhibit artefacts that are not completely removed, while the predictions from GUNet are much smoother and reproduce high contrast areas and edges with greater fidelity.

%% file: bigsample.tex
\begin{figure*}[t]%
\centering%
\begin{subfigure}{0.22\linewidth}%
\includegraphics[width=1.0\linewidth]{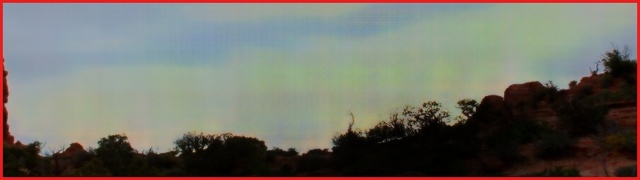}\\%
\includegraphics[width=1.0\linewidth]{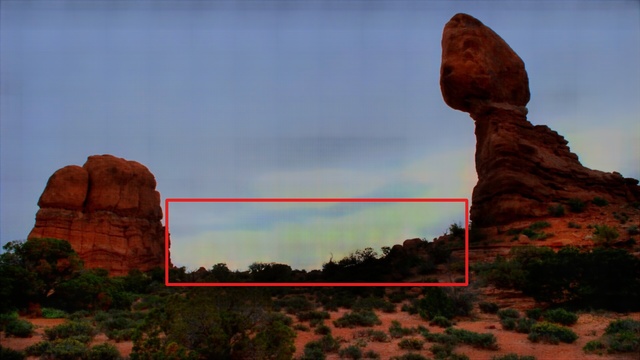}\\%
\includegraphics[width=1.0\linewidth]{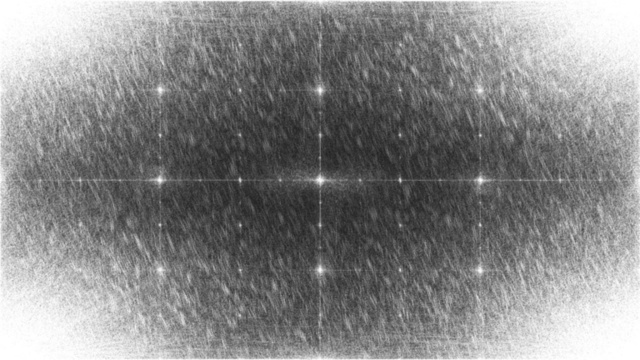}%
\caption{TC-UNet}%
\end{subfigure}%
\begin{subfigure}{0.22\linewidth}%
\includegraphics[width=1.0\linewidth]{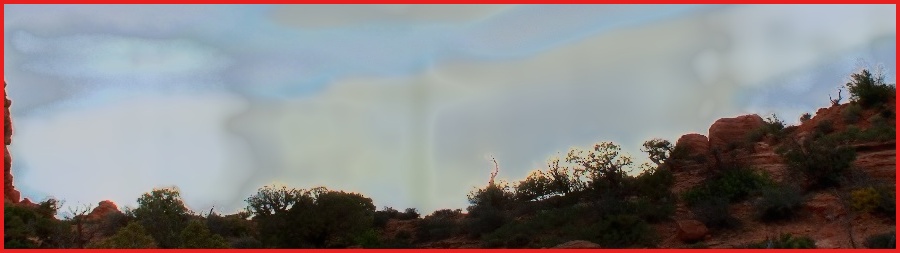}\\%
\includegraphics[width=1.0\linewidth]{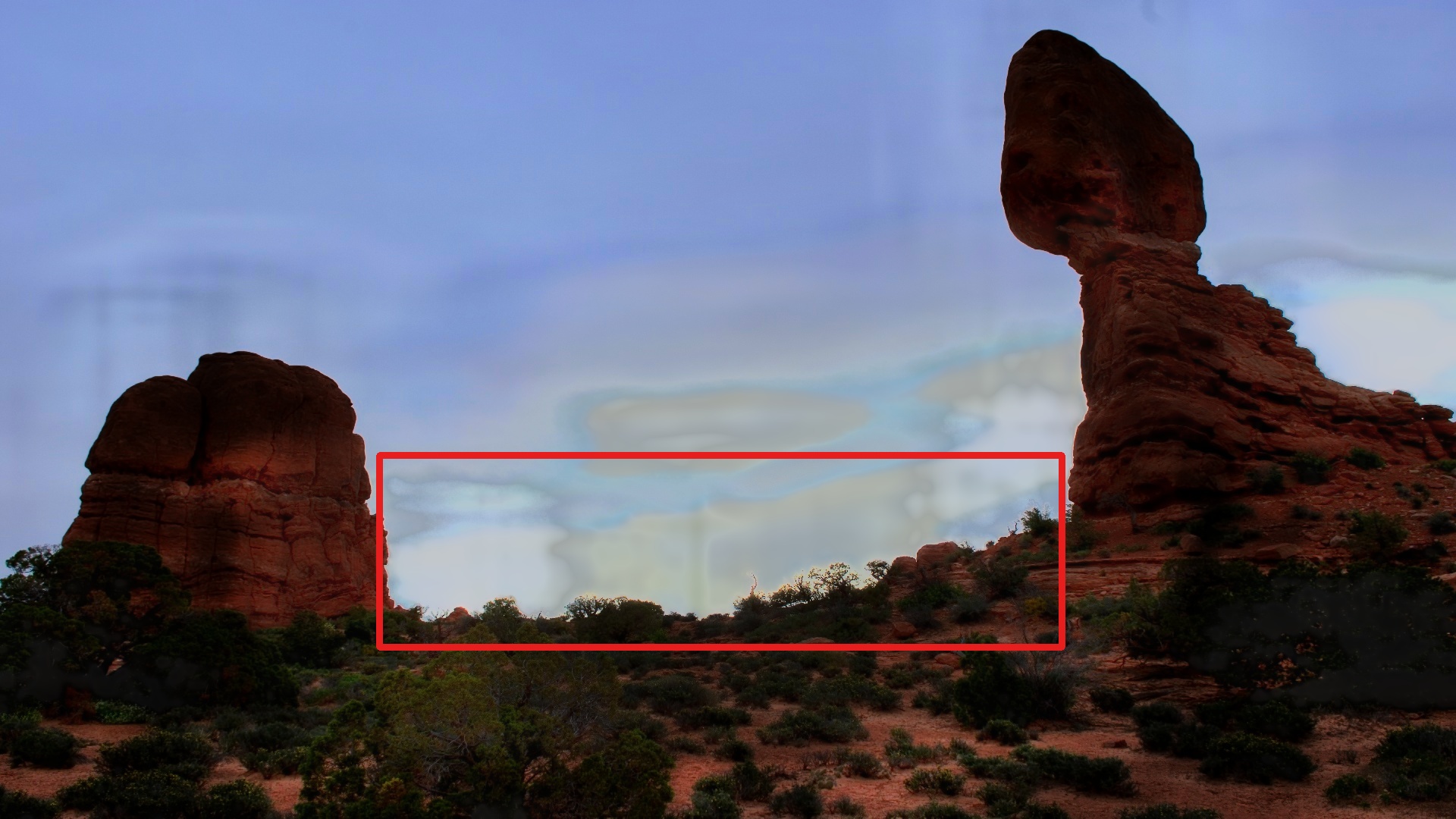}\\%
\includegraphics[width=1.0\linewidth]{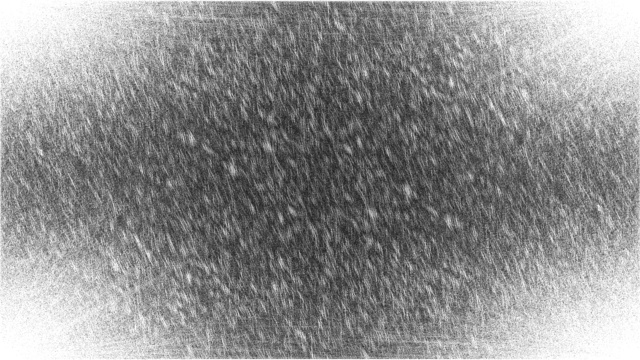}%
\caption{BI-UNet}%
\end{subfigure}%
\begin{subfigure}{0.22\linewidth}%
\includegraphics[width=1.0\linewidth]{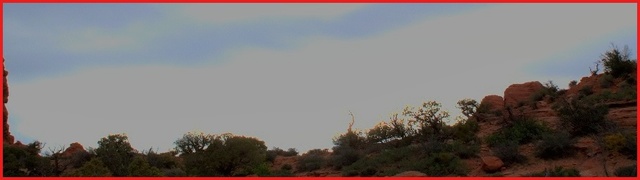}\\%
\includegraphics[width=1.0\linewidth]{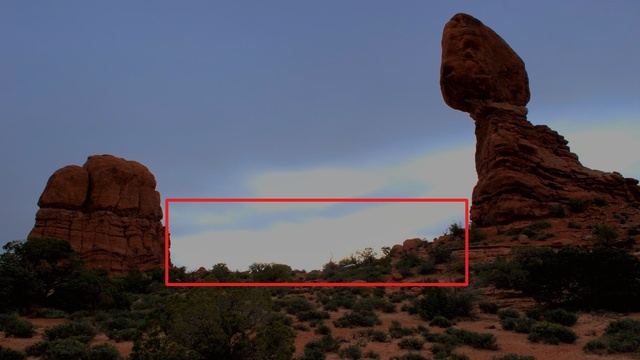}\\%
\includegraphics[width=1.0\linewidth]{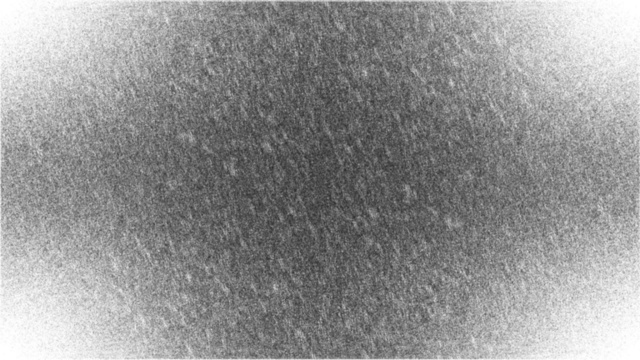}%
\caption{GUNet}%
\end{subfigure}%
\begin{subfigure}{0.22\linewidth}%
\includegraphics[width=1.0\linewidth]{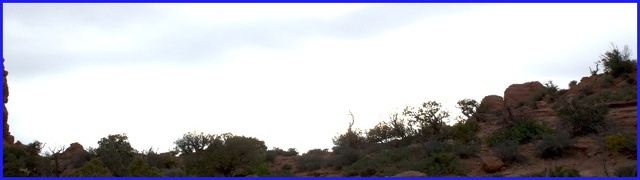}\\%
\includegraphics[width=1.0\linewidth]{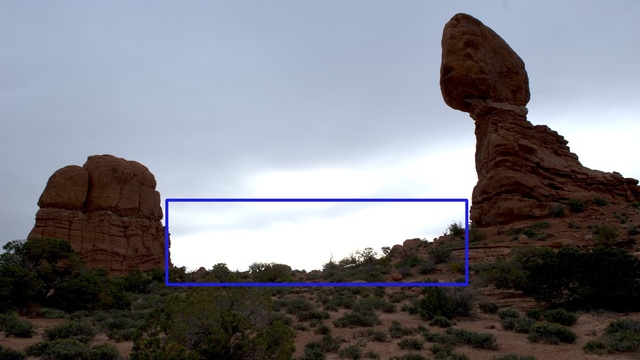}\\%
\includegraphics[width=1.0\linewidth]{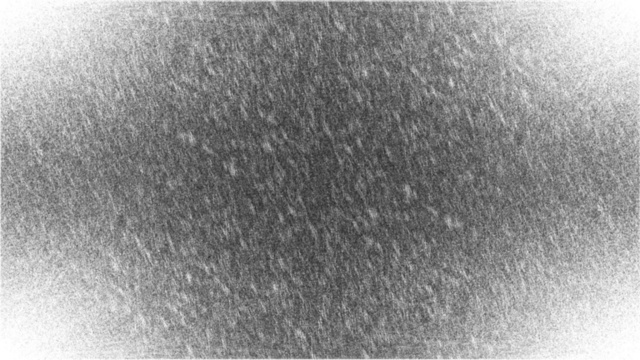}%
\caption{Input}%
\end{subfigure}%
\caption{Comparison of outputs from TC-UNet, BI-UNet and GUNet. The bottom row shows corresponding frequency spectra highlighting the induced artefacts from TC-UNet and a much better preserved spectrum from GUNet.}\label{fig:big_samples}
\end{figure*}

%% file: samples.tex
\begin{figure*}%
\centering%
\begin{subfigure}{0.49\linewidth}
\cfbox{myred}{\includegraphics[width=0.16\linewidth]{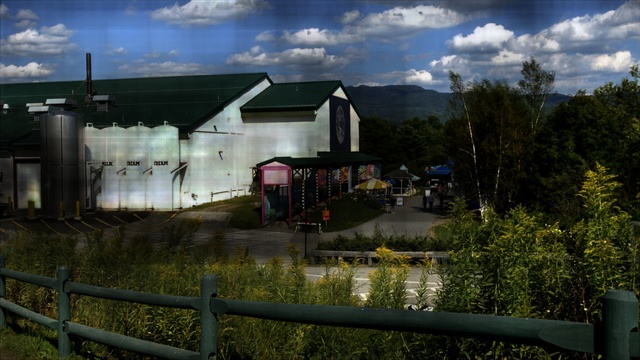}%
\includegraphics[width=0.16\linewidth]{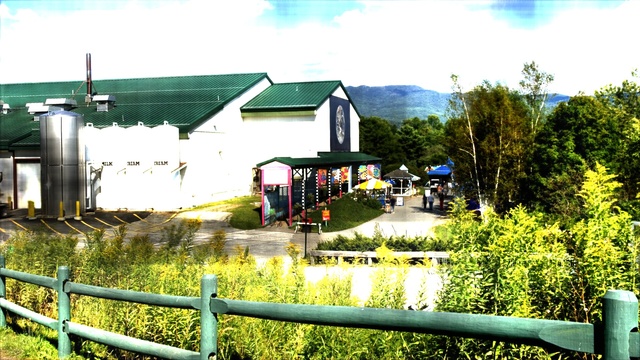}%
\includegraphics[width=0.16\linewidth]{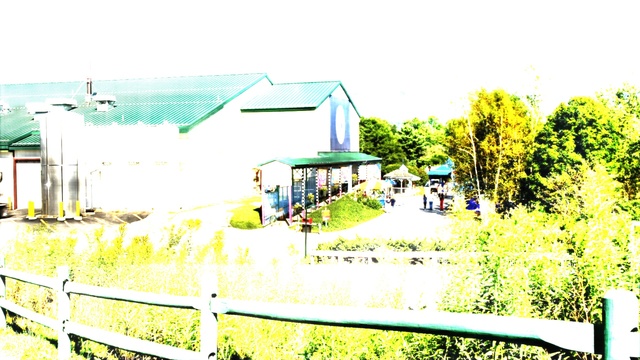}}%
\cfbox{mygreen}{\includegraphics[width=0.16\linewidth]{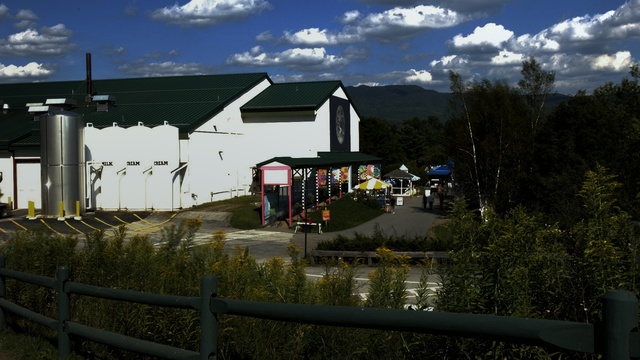}%
\includegraphics[width=0.16\linewidth]{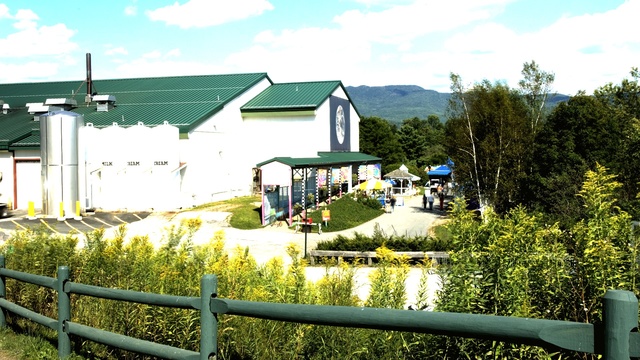}%
\includegraphics[width=0.16\linewidth]{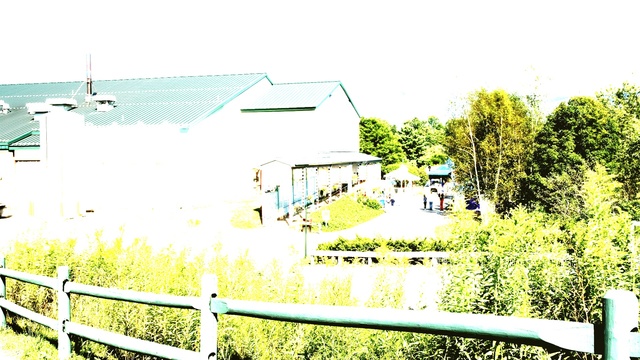}}\\%
\includegraphics[width=0.32\linewidth]{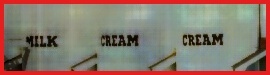}%
\includegraphics[width=0.32\linewidth]{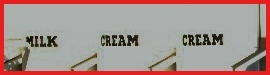}%
\includegraphics[width=0.32\linewidth]{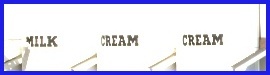}\\%
\includegraphics[width=0.32\linewidth]{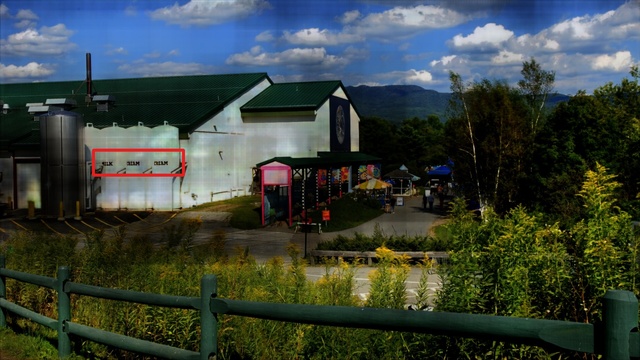}%
\includegraphics[width=0.32\linewidth]{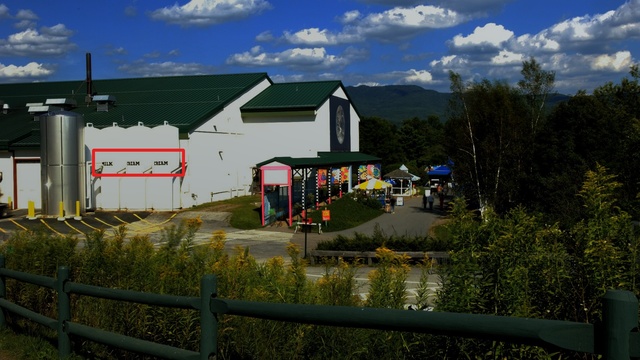}%
\includegraphics[width=0.32\linewidth]{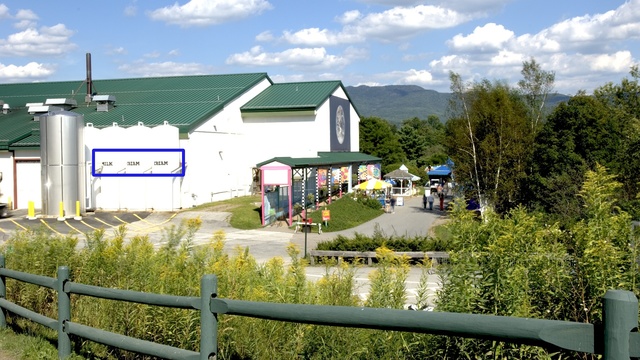}\\%
\includegraphics[width=0.32\linewidth]{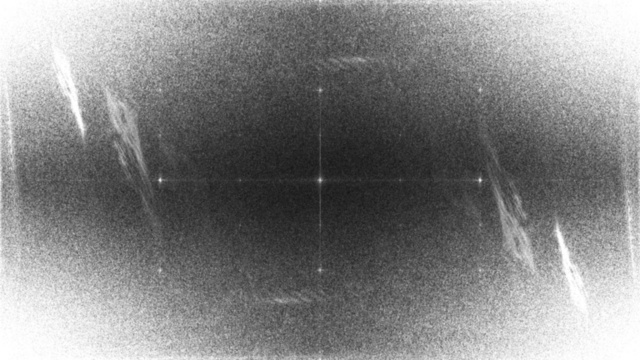}%
\includegraphics[width=0.32\linewidth]{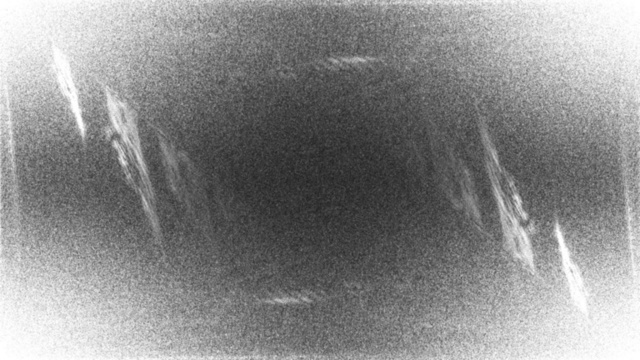}%
\includegraphics[width=0.32\linewidth]{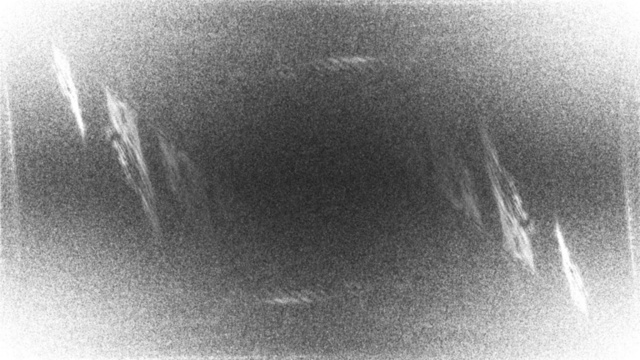}%
\caption{TC-UNet - GUNet - Input}
\end{subfigure}
\begin{subfigure}{0.49\linewidth}
\cfbox{myred}{\includegraphics[width=0.16\linewidth]{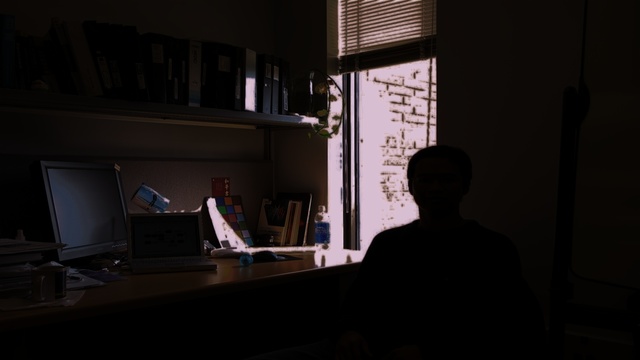}%
\includegraphics[width=0.16\linewidth]{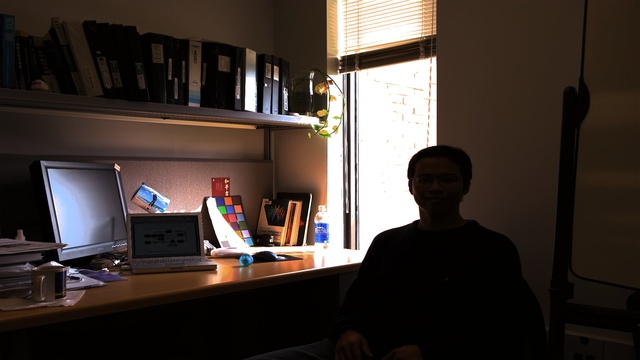}%
\includegraphics[width=0.16\linewidth]{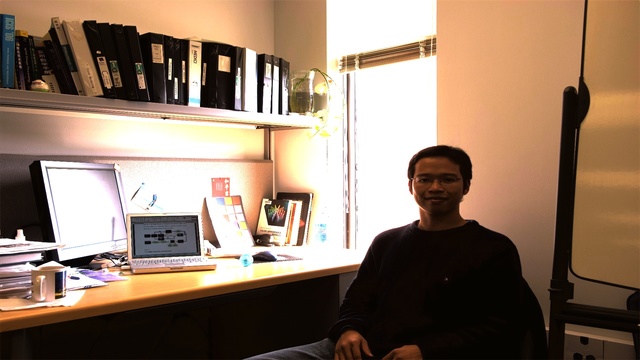}}%
\cfbox{mygreen}{\includegraphics[width=0.16\linewidth]{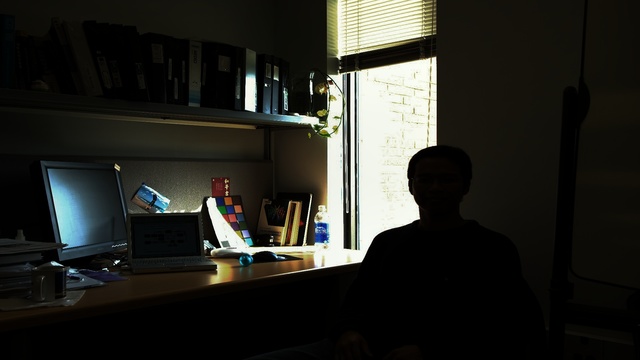}%
\includegraphics[width=0.16\linewidth]{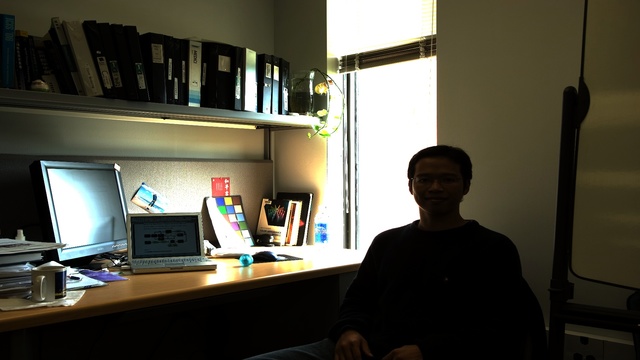}%
\includegraphics[width=0.16\linewidth]{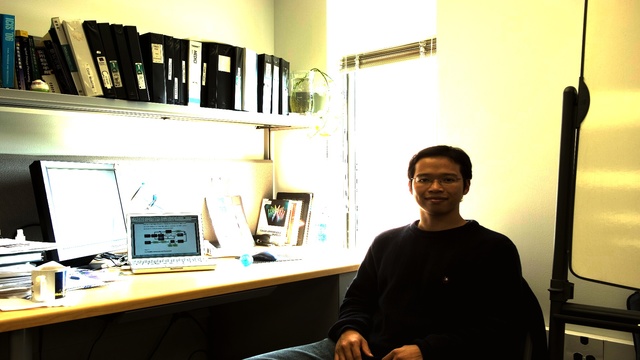}}\\%
\includegraphics[width=0.32\linewidth]{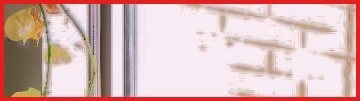}%
\includegraphics[width=0.32\linewidth]{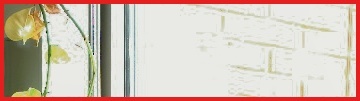}%
\includegraphics[width=0.32\linewidth]{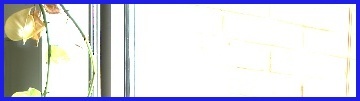}\\%
\includegraphics[width=0.32\linewidth]{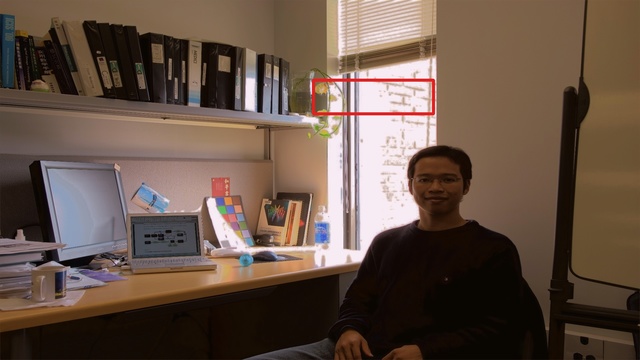}%
\includegraphics[width=0.32\linewidth]{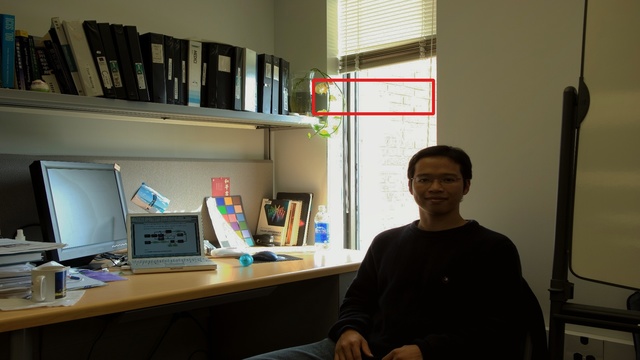}%
\includegraphics[width=0.32\linewidth]{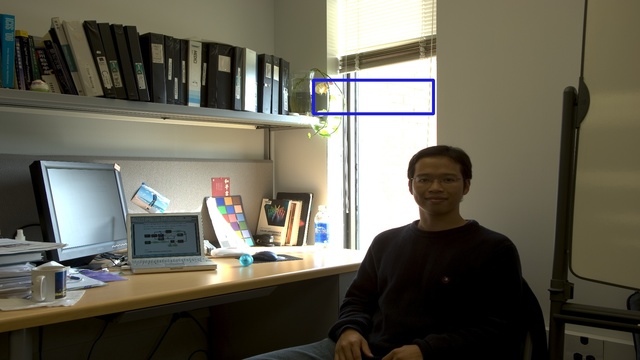}\\%
\includegraphics[width=0.32\linewidth]{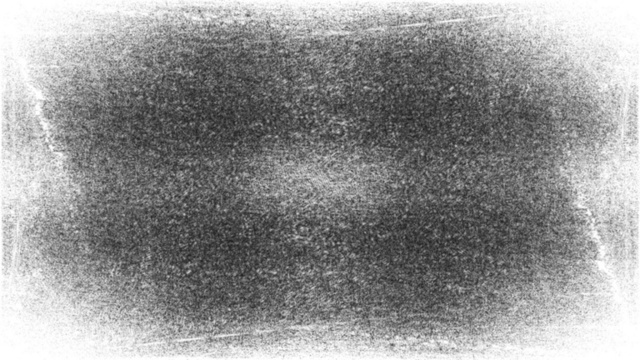}%
\includegraphics[width=0.32\linewidth]{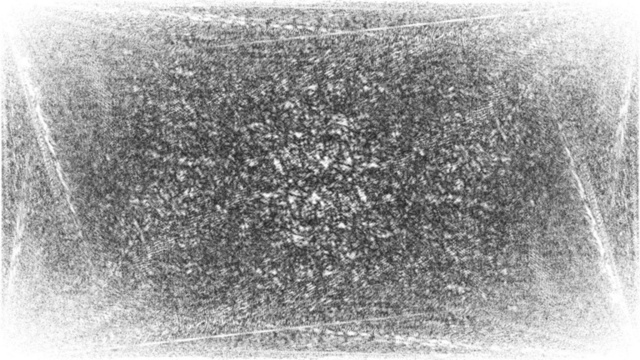}%
\includegraphics[width=0.32\linewidth]{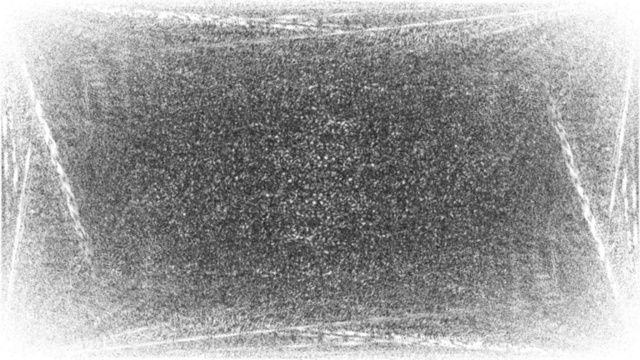}%
\caption{EXP - GUNet - Input}
\end{subfigure}
\caption{ITM predictions for TC-UNet, EXP and GUNet. (row 1) Output exposures. (row 2--3) Tone
mapped outputs. (row 4) Spectra. (b) is an example where a non-UNet based
CNN produces artefacts, which are
more pronounced in the spectrum.}\label{fig:samples}
\end{figure*}

%% file: tables.tex
\begin{table*}[h!]
\centering
\begin{tabular}{|l|c|c|c|c|c|c|}
\hline
\vrule width 0pt height 2.2ex
\multirow{3}{*}{Method}&\multicolumn{3}{c|}{\textit{scene-referred}}&\multicolumn{3}{c|}{\textit{display-referred}}\\\cline{2-7}
\vrule width 0pt height 2.2ex
&SSIM&MS-SSIM&HDR-VDP&SSIM&MS-SSIM&HDR-VDP\\\cline{2-7}
\vrule width 0pt height 2.2ex
&\textit{opt} / \textit{cull}&\textit{opt} / \textit{cull}&\textit{opt} / \textit{cull}&\textit{opt} / \textit{cull}&\textit{opt} / \textit{cull}&\textit{opt} / \textit{cull}\\\hline
\vrule width 0pt height 2.2ex
UNT&$0.68$ / $0.77$&$0.71$ / $0.70$&$34.9$ / $34.7$&$0.72$ / $0.78$&$0.73$ / $0.69$&$35.7$ / $35.3$\\
BIU&$0.79$ / $0.81$&$0.69$ / $0.65$&$37.2$ / $33.6$&$0.80$ / $0.81$&$0.70$ / $0.65$&$38.2$ / $34.4$\\
EIL&$0.72$ / $0.52$&$0.78$ / $0.53$&$39.0$ / $28.1$&$0.77$ / $0.54$&$0.80$ / $0.55$&$41.0$ / $27.6$\\
EXP&$0.74$ / $0.81$&$0.79$ / $\mathbf{0.79}$&$39.3$ / $\mathbf{35.0}$&$0.79$ / $0.83$&$0.82$ / $\mathbf{0.79}$&$40.8$ / $\mathbf{36.2}$\\
GUN&$\mathbf{0.84}$ / $\mathbf{0.84}$&$\mathbf{0.84}$ / $\mathbf{0.79}$&$\mathbf{40.6}$ / $33.4$&$\mathbf{0.85}$ / $\mathbf{0.84}$&$\mathbf{0.84}$ / $0.78$&$\mathbf{41.9}$ / $34.5$\\
\hline
\end{tabular}
\caption{Average values of the metrics for all methods.}\label{table:gunet_results}
\end{table*}

%% file: colour_figure.tex
\begin{figure*}[ht]
\centering
\includegraphics[width=0.9\linewidth]{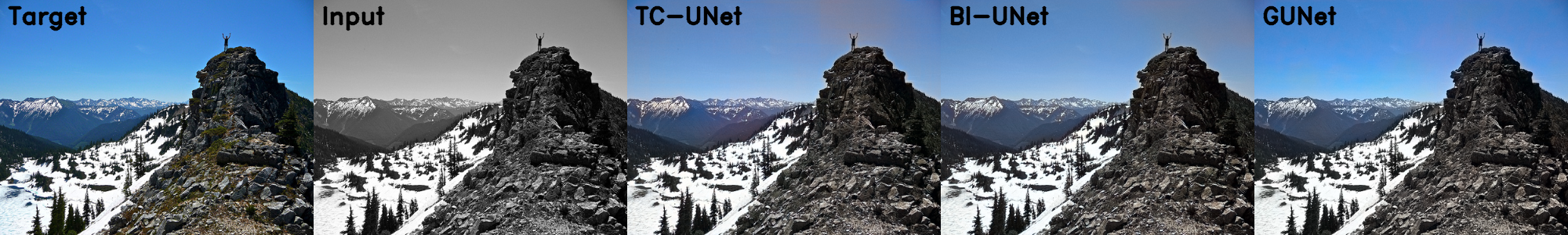}\\
\includegraphics[width=0.9\linewidth]{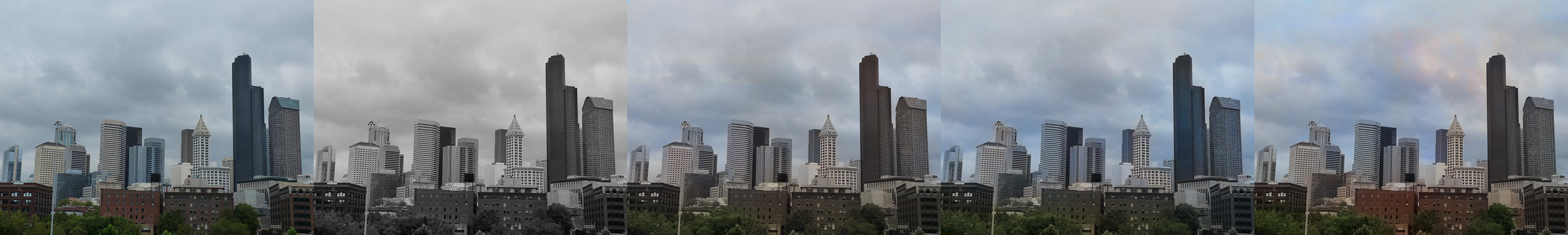}\\
\includegraphics[width=0.9\linewidth]{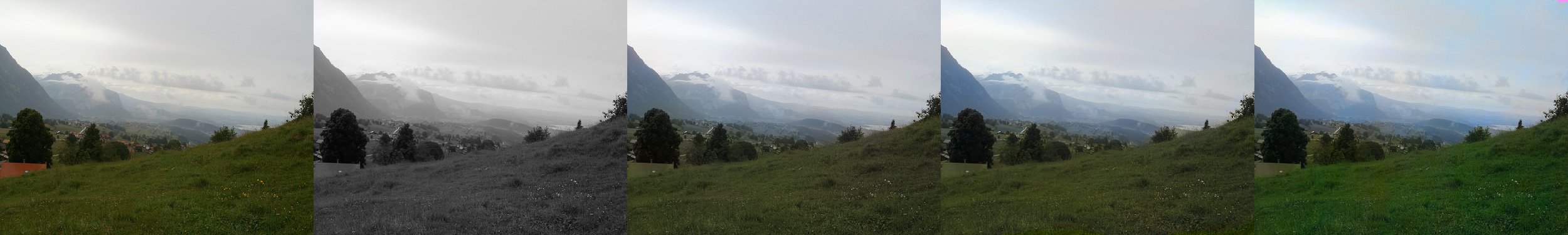}\\
\includegraphics[width=0.9\linewidth]{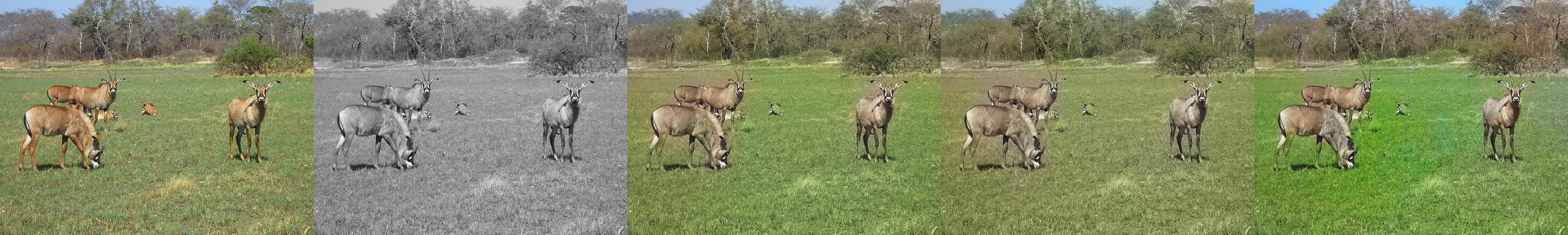}\\
\includegraphics[width=0.9\linewidth]{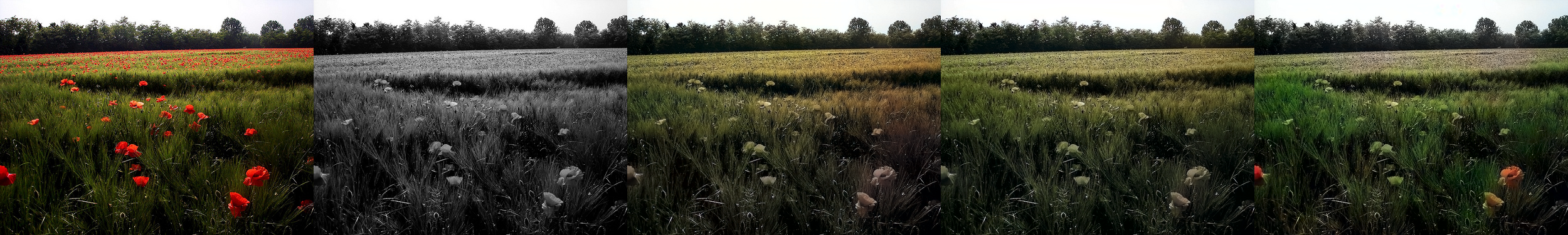}\\
\includegraphics[width=0.9\linewidth]{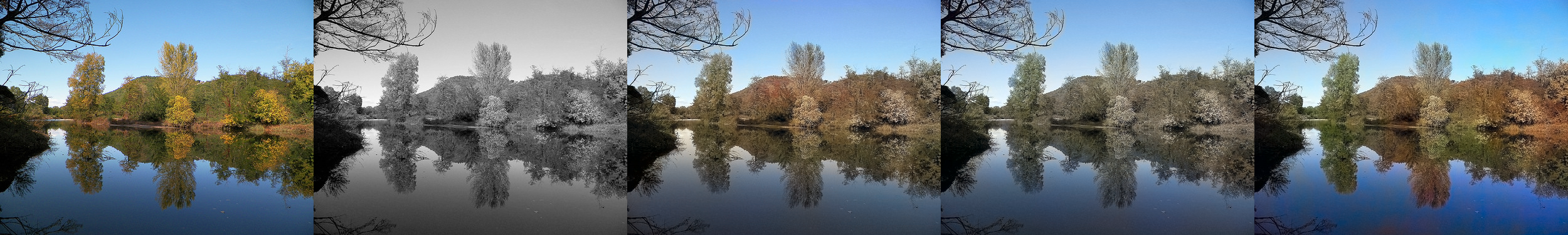}\\
\includegraphics[width=0.9\linewidth]{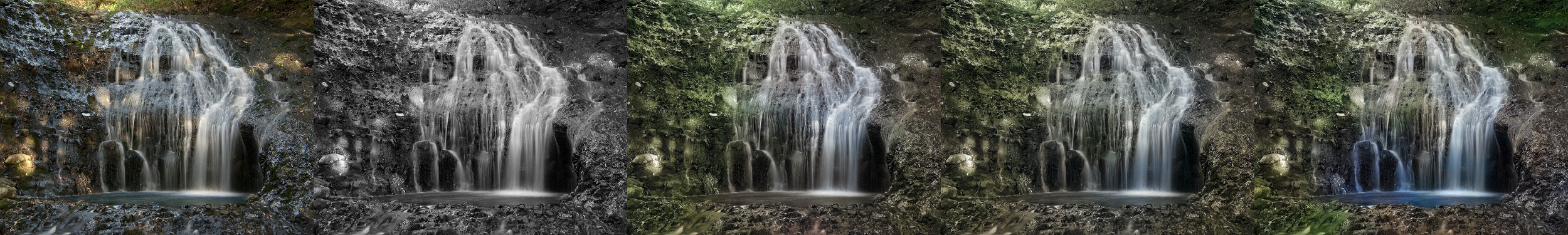}\\
\includegraphics[width=0.9\linewidth]{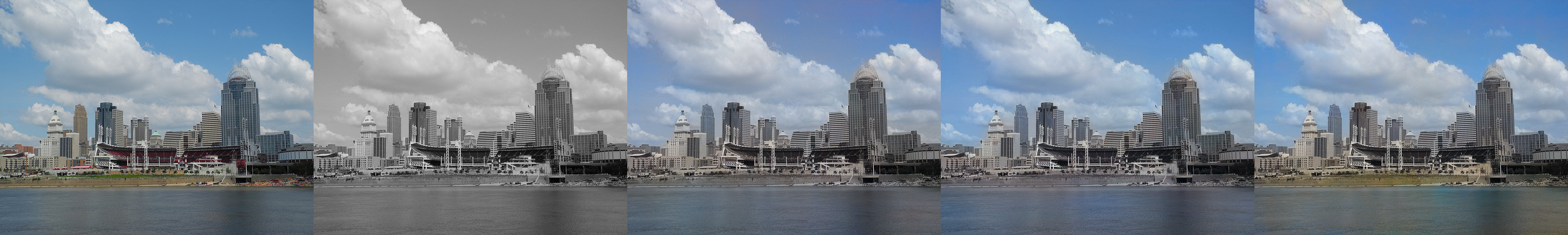}
\caption{Results for colourisation}\label{fig:colourfig}
\end{figure*}

%% file: colourisation.tex
\section{Application: Colourisation}\label{sec:colourisation}

Colourisation from greyscale images is another application that benefits from strong guidance from the input and from intermediate encoder features. This application is difficult to judge quantitatively since commonly used metrics, such as PSNR, do not account for multi-modality (e.g. alternative colourisation) and are also averaging, which can be misleading and overvalue desaturated/grey results. This section presents qualitative results for colourisation using GUNet, exhibiting the benefits from using a spectrally consistent architecture compared to other UNet alternatives.
To accelerate training, an ImageNet pre-trained 50-layer resnet with fixed weights is used as the encoder for all networks, leveraging knowledge transfer from a classifier. 
The decoders are composed of [conv2d - batchnorm - relu - conv2d] modules at each level, differing only at the fusion level. The guided fusion layer for GUNet uses a kernel size of 3 $\times$ 3 and $\epsilon = 0.001$. All networks are trained using a smooth L1 loss for 800,000 iterations using the Adam optimiser with a learning rate of 1e-3 and a batch size of $24$. The Places365~\cite{zhou2017places} dataset is used for training. A plot of the training loss is provided in Figure~\ref{fig:colour_train_loss}.

Figure~\ref{fig:colourfig} shows results for colourisation obtained using a TC-UNet, a BI-UNet and a GUNet architecture. The samples are from a test set composed of images collected from Flickr~\cite{flickr} The results highlight the benefit of using guidance in the architecture, as this minimises colour bleeding into surrounding objects. This can be observed in the sky around the person on the mountain, the clouds and tops of buildings and the reproduction of the grass in the cityscape. In addition, GUNet provides more colourful images, and can spatially adapt hues more quickly, providing higher local hue contrast.

\input{colourisation_loss}

%% file: colourisation_loss.tex
\begin{figure}[t!]
\centering
\includegraphics[width=0.5\linewidth]{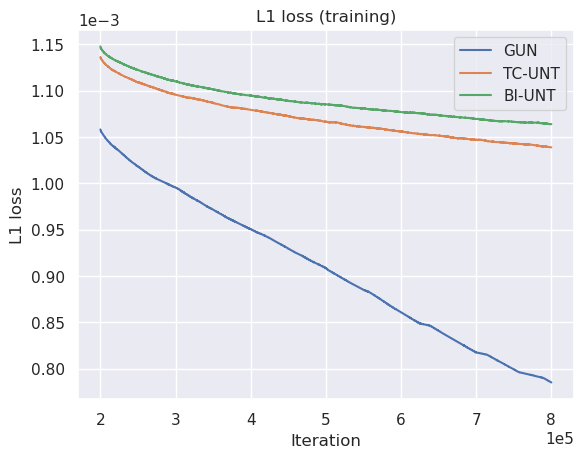}
    \caption{Training loss for colourisation.}\label{fig:colour_train_loss}
\end{figure}

%% file: conclusion.tex
\section{Conclusion}\label{sec:conclusion}

This work proposed GUNet, an architecture that improves the prediction quality of UNet-like architectures. This was achieved via the use of a novel upsampling module, based on guided image filtering. The effects that the structural biases of CNNs have on network outputs were investigated in the Fourier domain, showing the improvement attained by GUNet. The effectiveness of this approach was demonstrated in an example application of inverse tone mapping, where state-of-the-art performance was achieved and for colourisation, where GUNet exhibits benefits compared to alternative UNet architectures.

%% file: main.bbl
\begin{thebibliography}{10}

\bibitem{iizuka2016colornet}
Satoshi Iizuka, Edgar Simo-Serra, and Hiroshi Ishikawa.
\newblock {Let there be Color!: Joint End-to-end Learning of Global and Local
  Image Priors for Automatic Image Colorization with Simultaneous
  Classification}.
\newblock {\em ACM Transactions on Graphics (TOG)}, 35(4):110:1----110:11,
  2016.

\bibitem{sugawarasr}
Yusuke Sugawara, Sayaka Shiota, and Hitoshi Kiya.
\newblock {Super-Resolution using Convolutional Neural Networks without Any
  Checkerboard Artifacts}.
\newblock {\em IEEE International Conference on Image Processing (ICIP)}, pages
  66----70, jun 2018.

\bibitem{eilertsen2017cnn}
Gabriel Eilertsen, Joel Kronander, Gyorgy Denes, Rafa{\l}~K Mantiuk, and Jonas
  Unger.
\newblock {HDR image reconstruction from a single exposure using deep CNNs}.
\newblock {\em ACM Transactions on Graphics (TOG)}, 36(6):178, 2017.

\bibitem{ronneberger2015unet}
Olaf Ronneberger, Philipp Fischer, and Thomas Brox.
\newblock U-net: Convolutional networks for biomedical image segmentation.
\newblock In {\em International Conference on Medical image computing and
  computer-assisted intervention}, pages 234--241. Springer, 2015.

\bibitem{isola2016pix2pix}
Phillip Isola, Jun-Yan Zhu, Tinghui Zhou, and Alexei~A Efros.
\newblock Image-to-image translation with conditional adversarial networks.
\newblock In {\em Proceedings of the IEEE conference on computer vision and
  pattern recognition}, pages 1125--1134, 2017.

\bibitem{endo2017drtmo}
Yuki Endo, Yoshihiro Kanamori, and Jun Mitani.
\newblock Deep reverse tone mapping.
\newblock {\em ACM Trans. Graph.}, 36(6):177--1, 2017.

\bibitem{jin2017inverse}
Kyong~Hwan Jin, Michael~T. McCann, Emmanuel Froustey, and Michael Unser.
\newblock {Deep Convolutional Neural Network for Inverse Problems in Imaging}.
\newblock {\em IEEE Transactions on Image Processing}, 26(9):4509--4522, sep
  2017.

\bibitem{pix2pixsuppl}
Phillip Isola, Jun-Yan Zhu, Tinghui Zhou, and Alexei~A Efros.
\newblock {Image-to-Image Translation with Conditional Adversarial Nets.
  Supplementary Material}, 2016.

\bibitem{zhang2017suppl}
Jinsong Zhang and Jean-Fran{\c{c}}ois Lalonde.
\newblock {Learning High Dynamic Range from Outdoor Panoramas. Supplementary
  Material}, 2017.

\bibitem{marnerides2018exp}
Demetris Marnerides, Thomas Bashford-Rogers, Jonathan Hatchett, and Kurt
  Debattista.
\newblock {ExpandNet: A Deep Convolutional Neural Network for High Dynamic
  Range Expansion from Low Dynamic Range Content}.
\newblock {\em Computer Graphics Forum}, 37(2):37--49, may 2018.

\bibitem{odena2016deconvolution}
Augustus Odena, Vincent Dumoulin, and Chris Olah.
\newblock Deconvolution and checkerboard artifacts.
\newblock {\em Distill}, 1(10):e3, 2016.

\bibitem{hegif}
Kaiming He, Jian Sun, and Xiaoou Tang.
\newblock {Guided Image Filtering}.
\newblock {\em IEEE Transactions on Pattern Analysis and Machine Intelligence},
  35(6):1397--1409, jun 2013.

\bibitem{banterle06itm}
Francesco Banterle, Patrick Ledda, Kurt Debattista, and Alan Chalmers.
\newblock {Inverse tone mapping}.
\newblock {\em Proceedings of GRAPHITE '06}, page 349, 2006.

\bibitem{schmidhuber2014deep}
J{\"{u}}rgen Schmidhuber.
\newblock {Deep Learning in neural networks: An overview}.
\newblock {\em Neural Networks}, 61:85--117, 2015.

\bibitem{longfullyconv}
Jonathan Long, Evan Shelhamer, and Trevor Darrell.
\newblock {Fully convolutional networks for semantic segmentation}.
\newblock {\em Proceedings of the IEEE Computer Society Conference on Computer
  Vision and Pattern Recognition}, 07-12-June:3431--3440, 2015.

\bibitem{shi2016real}
Wenzhe Shi, Jose Caballero, Ferenc Husz{\'a}r, Johannes Totz, Andrew~P Aitken,
  Rob Bishop, Daniel Rueckert, and Zehan Wang.
\newblock Real-time single image and video super-resolution using an efficient
  sub-pixel convolutional neural network.
\newblock In {\em Proceedings of the IEEE conference on computer vision and
  pattern recognition}, pages 1874--1883, 2016.

\bibitem{wojna2017devil}
Zbigniew Wojna, Jasper~RR Uijlings, Sergio Guadarrama, Nathan Silberman,
  Liang-Chieh Chen, Alireza Fathi, and Vittorio Ferrari.
\newblock The devil is in the decoder.
\newblock In {\em BMVC}, 2017.

\bibitem{aitken2017checkerboard}
Andrew Aitken, Christian Ledig, Lucas Theis, Jose Caballero, Zehan Wang, and
  Wenzhe Shi.
\newblock Checkerboard artifact free sub-pixel convolution: A note on sub-pixel
  convolution, resize convolution and convolution resize, 2017.

\bibitem{kopf2007joint}
Johannes Kopf, Michael~F Cohen, Dani Lischinski, and Matt Uyttendaele.
\newblock {Joint bilateral upsampling}.
\newblock In {\em ACM Transactions on Graphics (ToG)}, volume~26, page~96. ACM,
  2007.

\bibitem{hefastgif}
Kaiming He and Jian Sun.
\newblock Fast guided filter.
\newblock {\em arXiv preprint arXiv:1505.00996}, 2015.

\bibitem{wutrainablegif}
Huikai Wu, Shuai Zheng, Junge Zhang, and Kaiqi Huang.
\newblock Fast end-to-end trainable guided filter.
\newblock In {\em Proceedings of the IEEE Conference on Computer Vision and
  Pattern Recognition}, pages 1838--1847, 2018.

\bibitem{heidentityresnet}
Kaiming He, Xiangyu Zhang, Shaoqing Ren, and Jian Sun.
\newblock Identity mappings in deep residual networks.
\newblock In {\em European conference on computer vision}, pages 630--645.
  Springer, 2016.

\bibitem{he2015delving}
Kaiming He, Xiangyu Zhang, Shaoqing Ren, and Jian Sun.
\newblock Delving deep into rectifiers: Surpassing human-level performance on
  imagenet classification.
\newblock In {\em Proceedings of the IEEE international conference on computer
  vision}, pages 1026--1034, 2015.

\bibitem{akyuz2007hdr}
Ahmet~Oǧuz Aky{\"{u}}z, Roland Fleming, Bernhard~E Riecke, Erik Reinhard, and
  Heinrich~H B{\"{u}}lthoff.
\newblock {Do HDR displays support LDR content?: a psychophysical evaluation}.
\newblock {\em ACM Transactions on Graphics (TOG)}, 26(3):38, 2007.

\bibitem{fairchild2007hdr}
Mark~D Fairchild.
\newblock The hdr photographic survey.
\newblock In {\em Color and imaging conference}, pages 233--238. Society for
  Imaging Science and Technology, 2007.

\bibitem{pytorch}
Adam Paszke, Sam Gross, Francisco Massa, Adam Lerer, James Bradbury, Gregory
  Chanan, Trevor Killeen, Zeming Lin, Natalia Gimelshein, Luca Antiga, Alban
  Desmaison, Andreas Kopf, Edward Yang, Zachary DeVito, Martin Raison, Alykhan
  Tejani, Sasank Chilamkurthy, Benoit Steiner, Lu~Fang, Junjie Bai, and Soumith
  Chintala.
\newblock Pytorch: An imperative style, high-performance deep learning library.
\newblock In H.~Wallach, H.~Larochelle, A.~Beygelzimer, F.~d\' Alch\'{e}-Buc,
  E.~Fox, and R.~Garnett, editors, {\em Advances in Neural Information
  Processing Systems 32}, pages 8024--8035. Curran Associates, Inc., 2019.

\bibitem{aydin2008extending}
Tun{\c{c}}~O Ayd{\i}n, Rafal Mantiuk, and Hans-Peter Seidel.
\newblock Extending quality metrics to full luminance range images.
\newblock In {\em Human Vision and Electronic Imaging XIII}, volume 6806, page
  68060B. International Society for Optics and Photonics, 2008.

\bibitem{wang2004image}
Zhou Wang, Alan~C Bovik, Hamid~R Sheikh, and Eero~P Simoncelli.
\newblock Image quality assessment: from error visibility to structural
  similarity.
\newblock {\em IEEE transactions on image processing}, 13(4):600--612, 2004.

\bibitem{wang2003multiscale}
Zhou Wang, Eero~P Simoncelli, and Alan~C Bovik.
\newblock Multiscale structural similarity for image quality assessment.
\newblock In {\em The Thrity-Seventh Asilomar Conference on Signals, Systems \&
  Computers, 2003}, volume~2, pages 1398--1402. Ieee, 2003.

\bibitem{narwaria2015hdr}
Manish Narwaria, Rafal Mantiuk, Mattheiu~P Da~Silva, and Patrick Le~Callet.
\newblock Hdr-vdp-2.2: a calibrated method for objective quality prediction of
  high-dynamic range and standard images.
\newblock {\em Journal of Electronic Imaging}, 24(1):010501, 2015.

\bibitem{zhou2017places}
Bolei Zhou, Agata Lapedriza, Aditya Khosla, Aude Oliva, and Antonio Torralba.
\newblock Places: A 10 million image database for scene recognition.
\newblock {\em IEEE Transactions on Pattern Analysis and Machine Intelligence},
  2017.

\bibitem{flickr}
Flickr.
\newblock {No Title}, 2018.

\end{thebibliography}
